# POWERS OF THE GOVERNMENTAL STATE
# AS FEEDBACK CONTROL DYNAMIC SYSTEM

Hokky Situngkir
(quicchote@yahoo.com)
Bandung Fe Institute

**Abstract**
Democracy in a state is possibly a gain to have the government that represents the whole citizens. But the main element of the democratic system is the ability of the governmental functions in the state to act properly in order to gain the citizen wants. The simulation presented in this paper shows how the rule-making, rule-application, and rule-adjudication functions work such ways and how they work to have the public wants. The model is the second order feedback control dynamic system where the rule-making and rule-application function placed in the forward path and the rule-adjudication placed at backward path of the system. The rule-making function assumed to be the function that accentuate the result of public inquiries and presented as a gain function. The rule-application and the rule-adjudication are modeled as an exponential function which response is lowering the entropy of the disordered state conditions by the policies that output. The possible noise that came from the inability of the government system to work properly is also presented here. The government system that succeed the gain the same as the public wants is the optimum government that wanted to be achieved.

Today if we are talking about governmental issues and the institutions that constitute the government, we will never be able to avoid using the terms of democracy. Democracy has been inherent issue to be raised on the social and political interests. Is it true that "democracy" has become the final, ultimate, and highest culmination of the evolution of human civilization (Fukuyama:1992)? Democracy, liberalization, capitalism have been the properties of the 21$^{st}$ century and is it true that what we can do now is to polish the basic concept of democracy and liberalization to cope the local conditions in some pre-democratic countries? This has been the basic conception that established the spirit of oxidentalism that raised among the injection of orientalism perspectives within the globalization (Turner:1994). Democracy, liberalization have been too often being used for other needs of vested interests that always lack of justice that shall be gained through democracy practically. This paper will not try to answer the question above, but to make other perspectives on how the main political powers in a nation-state shall be organized in order to gain justice.

## 1. Basic Concept of Powers and State in Democracy

The main background to birth the discourse of democracy was to avoid despotic nation-state, tyrannical government that oppresses the citizens. According to Montesquieu, there are three category of government, i.e.: the republican government (subdivided into democratic and aristocratic government), monarchic and despotic government (Montesquieu:1752 ch.2). We will use the main concept of this separation of power in this paper with assumption that all of the states of government exist will always can be divided into these three categories of power. In practice, the government of states will always become the variety to the three kinds of power, but the implementation will not always into three institutions. Some of the states combine two or even three of the category of powers, but some other made some variation for example, separating one of the category become two or even more institutions in government. These facts depend upon the conditions of the state spatio-temporally and the situational arguments that raised beyond the needs of the state.

The legislative power is the governmental power that enacts temporary or perpetual laws and amends or abrogates those that have been already enacted (Montesuiquieu:1752, ch.5). Moreover, the legislative will direct how the whole government shall be employed for preserving the community the whole citizens (Locke:1688 ch.12). The ideal condition for the legislative power is that the whole citizens should be seated in legislative power. But it is almost impossible to do that. Therefore, moderately the representations of the whole people shall be seated in the legislative power with some mechanisms to be made where the representatives decree their responsibility for the people their represented (Gunning:2000). In other words, the public citizen shall transact by their representatives what they cannot transact by themselves (Montesquieu:1752 ch.5).

The executive power is the governmental power that makes peace of war, leagues or alliances, sends or receives embassies, establishes the public security and provides against invasions, and all transactions with all persons and communities as the representative of the state (Montesquieu:1752, ch.5, Locke:1688, ch.12). The executive power also shall be represented to a person, not too many of persons since the policies in this power shall be made need short of time to be made. Other reasons is because the executive power often need a deep comprehension for the policies, so that there must be a free situation for the executive power to act cooperation with the experts and the progress will be seen whether to be accepted or rejected by the legislative power.

The judicial power is the power for punishment of the criminals, determines the disputes that arises between individuals (Montesquieu:1752, ch.5). In other words, the judicial power has the duty of arbitration, pronounces on special case and not upon general principles, and has the property of inability to act unless it is appealed to or until it has taken cognizance of an affair (Tocqueville:1840, ch.6).

The three kinds of power will share the unity of the whole governmental power in a state. As once stated by John Locke, democracy has an important property to do with those powers, namely the separation of those three kinds of power. Associations or coalitions among those powers will tend to the condition of anti-democratic or even despotic, because:

> the laws, that are at once, and in a short time made, have a constant and lasting force, and need a perpetual execution, or an attendance thereunto; therefore it is necessary there should be a power always in being, which should see to the execution of the laws that are made, and remain in force. And thus the legislative and executive power come often to be separated. (Locke:1688, ch.12).

Thus, Montesquieu even more deeply stated:

> there is no liberty, if the judiciary power be not separated from the legislative and executive. Were it joined with the legislative, the life and liberty of the subject would be exposed to arbitrary control; for the judge would be then the legislator. Were it joined to the executive power, the judge might behave with violence and oppression... (Motensquieu:1752, ch.11).

Henceforth, the three governmental powers shall never implemented in one hand of powers unless the state will fall into the despotism. All of this rule on the government implemented theoretically in the constitution of the state as the basic social contract between the citizen and the government. Therefore the constitution become the highest law in such a nation, and everything concerning the governmental system shall be realized by the constitution, otherwise the constitution is not complete and it will be the leak that to be the hole for despotic or even dictatorship government (Russell:2000).

By this basic definitions we can obviously see that in the discourse of the rule of law, we simply say that in governmental institutions can be classified in three kinds of functions i.e.: rulemaking function that represented by the legislative power, rule-application function that represented by the executive power, and the third, rule-adjudication function that taken by the judicial power (Bloch:1986). In this paper and so forth we will use these terms, as the model we present here we will made in function of law-production.

In advance, we will use this basic of thinking to build up a model of the government, and we will see how the institutionalized powers interacting each other to run the government office in mathematical model.

**2. Building The Model**

We build the model based on the fact that the three governmental functions interact among them in the rule of law perspective. Therefore, we will see how the three functions concerning to their action in the law perspective. We will try to model the action of each institution in their effort to achieve their each aim in the perspective of the law abiding.

The rule making function, as the legislative institution, can be seen as the function to make translations of what the citizen wants to the law-language. In other words, the rule making function is to accentuate the aspiration to the words of law. Here the duty of the legislative will depend on the public inquiry in order to have the simple points of what the people wants. By this, we can say that the response function of the legislative as a gain function:

$$D_L(t) = \alpha,$$

where $D_L$ is the entropy of what people wants and $\alpha$ is the coefficient that represents the ability to accentuate the citizen's aspiration; the bigger $\alpha$ the more public aspiration can be translated into the rule of governance.

In other hand, the rule application function can be seen as the ability of the executive institution to interpret the rule of governance made by the rule-maker. Thus, the rule application can be modeled as the function for the application and implementation of the rule that made by the legislative parts. In this case, the application can be seen as execution of policies in the citizen area – how to order and manage the state from the big entropy into the lowest entropy as well as stated by the rule-makers. Here, we can picture this function as an exponential function,

$$D_E(t) = \exp(-\beta t),$$

where $D_E$ is the response of the rule-application function that try to lower the entropy of what people wants in executive or application manner, and $\beta$ is the coefficient of the acceleration of the rule application to be established. It is obvious that

$$\lim_{t \to \infty} D_E(t) = 0$$

and the bigger the β the faster the functions accomplished.

The function of rule adjudication can be seen also in this view. The function is to become the arbitrary function on any disputes. Thus, this function is to lower the entropy of the condition of the system. Here, we can make up

$$D_J(t) = \exp(-\gamma t),$$

Where $D_J$ is the response of the adjudication function. And so forth, γ is the coefficient of the function to be accelerated. While,

$$\lim_{t \to \infty} D_J(t) = 0$$

and the bigger the γ the faster the functions accomplished.

We can see obviously how this components establish the dynamic feedback control system. Figure 1 shows the outlined model of the relation that established the dynamic feedback control. By the figure we can see that we have some signals of law that apprehended by the functions of each elements. The signals can be describe below,

$P_{in}$ ≡ process of inquiring to people wants. The process here can be done by the elections, polling, referendum, mass-media, public-demonstrations, intelligence services, et cetera. In short, this is the outcome of the whole process to inquiry what people want the government to do.

$P_L$ ≡ laws enacting according with considerations on what people want and some input from the arbitration process. The output procedures can be:
✓ making a new law/rule
✓ amending the old one rules/laws
✓ abrogating the old one rules/laws.
The whole data from the inquiring process must be digested well by the representatives in order to have laws to be reinforced reflecting the needs and the people want.

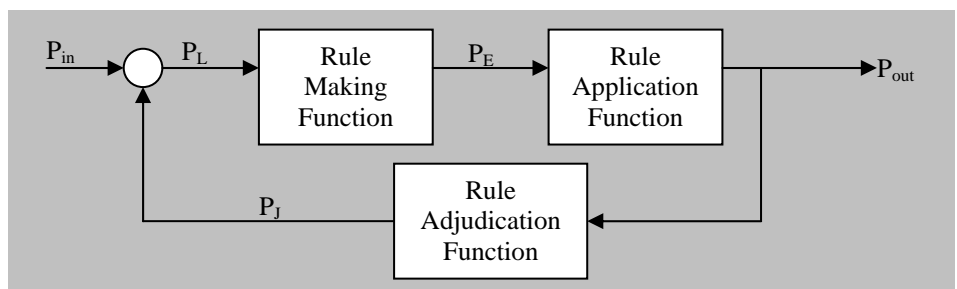

**Figure 1**
Modified feedback control as a model of governmental powers

$P_E$ ≡ laws to be applied into public policies. This is the rule-making function of legislative body. They also become the input or the main guidance for the executive in their duties as rule-application function of government.

$P_{out}$ ≡ public policies concerning governmental issues and state's resources. This is the policies and the central management process of a state or a country. The face of the country will be mostly affected by the policies that enacted by the executive.

Formally, the analytical functions refer to each power can be stated as:
- Legislative process : $P_L \rightarrow P_E + X_L$
- Executive process : $P_E \rightarrow P_{out} + X_E$
- Judicial process : $X_L + X_E \rightarrow P_J$

Or mathematically:

$$D_L(s) \approx \frac{P_E(s)}{P_L(s)}, \quad D_E(s) \approx \frac{P_{out}(s)}{P_E(s)} \quad \text{and} \quad D_J(s) \approx \frac{P_J(s)}{P_{out}(s)},$$

while $D_L(s)$, $D_E(s)$, and $D_J(s)$ are $D_L(t)$, $D_E(t)$, $D_J(t)$ respectively after the Laplace transform.

The $P_L$ can also be written as $(P_{in} - P_J)$, as the input for the legislation process must concern to the inquiring public needs and wants but also the arbitration on possibility of conflicts amongst the whole citizens that soluted by the judicial power of the state.

$X_L$ and $X_E$ are some uncoped aspects that possibly comes out from the inefficiency of the body, some noise that can be apprehend as disturbances on the process of legislation or executions. These aspects mostly become the sources of disputes or conflicts amongst citizens, amongst the representatives in legislative body, citizens (possibly NGOs as extra-parliamentary power) to the executive body, or even the whole governmental system.

By now then, we can have the feedback control system of the governmental system as described in figure 2. In advance we will use this model the conditions that may appear and how the response of the system simulated to cope such a condition.

At a glance, it is obvious that our model turns out to be the simple 2$^{rd}$ order control system. We can see that the transfer function of the whole system turns to be:

$$Transfer\_Function = \frac{\alpha}{(\beta + s)\left(1 + \frac{\alpha}{(\beta + s)(s + \gamma)}\right)}$$

where each α, β and γ shall be made upon some degree of each elements to work out.

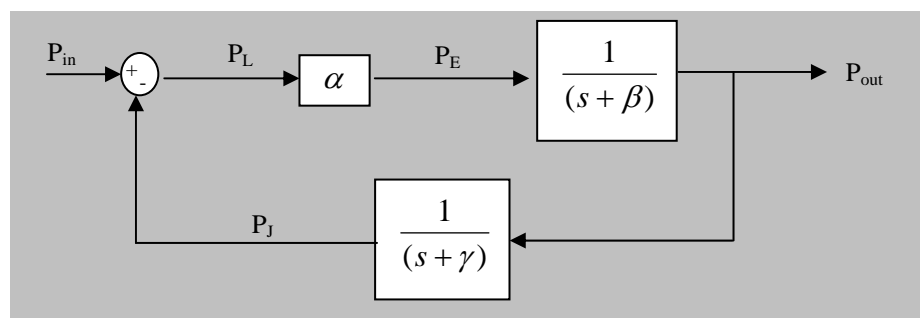

**Figure 2**
The 2$^{nd}$ order dynamical feedback control system that represents the interactions among powers in a state.

The good combination among the values of α, β and γ, will bring the system to the more effective of the outputs of the governmental system. We must remember here that even while the α, β, and γ is very big that represented each elements works as fast as possible, the system will not even give the expected results. The way of our simulation will gain the answer key on this combination.

### 3. Simulation Results

In this section, we will analyze the response of the whole system with some conditions that will reveal computationally some aspects of the governmental system that usually has been taken for granted by the qualitative analysis, and become the evolution of democratic system that we have now. This analysis will be also hoped to make us the ability to measure the effectiveness of some governmental system that we have in reality.

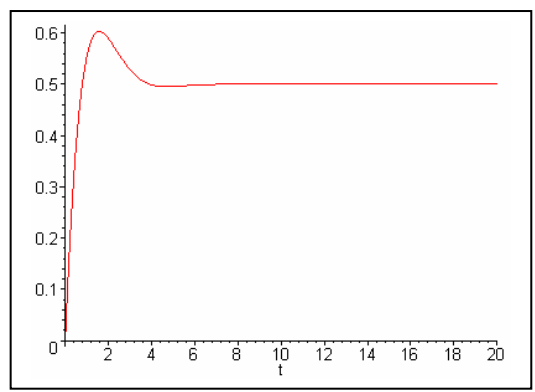

**Figure 3**

The response of governmental system while α=β=γ=1. The y-axis is the state condition of policies while the x-axis is the time. We can see that while the input of what people wants is in the state of 1 (step function), the output policies is by only 0.5 in the steady state condition. Look that in the first 4 time-sequence the system is attracted to be oscillation with the maximum overshot is 0.6. We must be presumed that what pictured here is a one constant round game and the system will be in the steady-state while the state of conditions are constant.

For the first example, we will have a kind of condition while the value of α=β=γ=1. Figure 3 shows the response system for this situation. In this case, we can see that while the input of what people wants is in the state of 1 (step function), the output policies is by only 0.5 in the steady state condition. Moreover, in the first 4 time-sequence the system is attracted to be oscillation with the maximum overshot is 0.6.

This example reflects that the governmental interactions are not in maximum optimization henceforth, the step-response we propose here is the situation of one-round game and while the system is in the constant state of conditions we will see the state-response that figured. However, the system should be evolved dynamically every round of game and the system should be more dynamical in practice until the steady state output gets the same state-value with what the citizen wants.

Now we will see how the system of governance will response while the γ=β=1 and varied values of α. Figure 4a showed how the system of government will be more and more optimum the bigger the value of α. While α=10, the system will be in under-damped dynamical control system wile in the first 6 time-sequence the response oscillated around the fixed-points of 1, while the same value of the step-response the people want it to be. By this fact, we can somehow say that the system will in more effective way while the value of α=10, β=γ=1. The more high value of α, the more effective the way system of governance goes. In reverse, the lower value of α, the more the system damped, and by this conditions we will hope that the elements of the system of governance will co-evolve such a way until the expected value of the output can be reached. Concerning the meaning of α as the coefficient that reflected whether the rule made in appropriate acceleration or not thus, qualitatively we can say that the system of governance will be much more efficient while the legislative can make the translation of what the public wants faster.

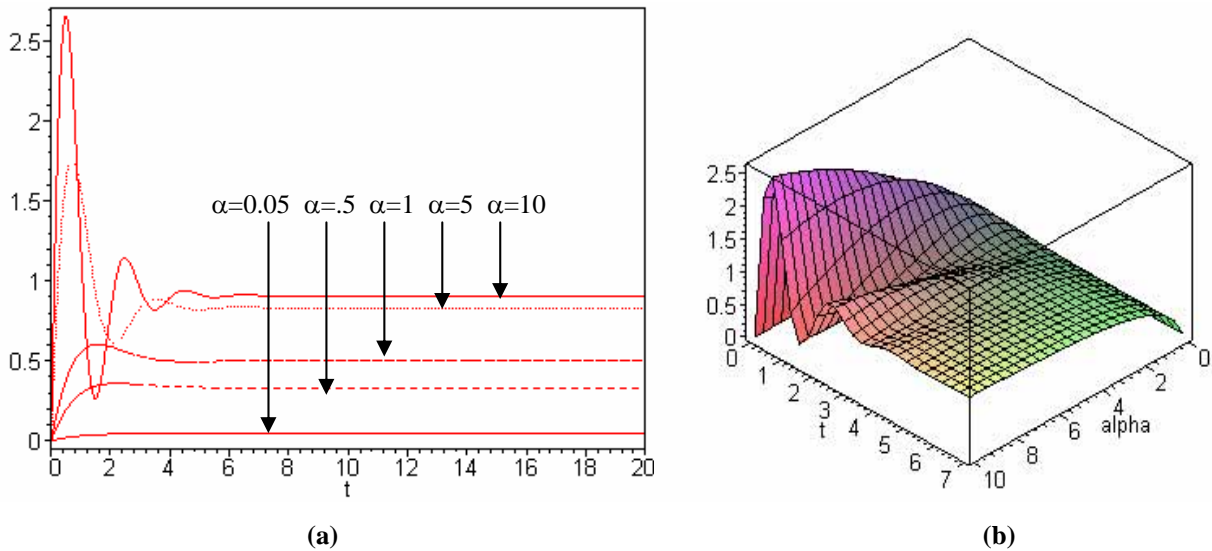

(a)    (b)

**Figure 4**

The more value of α - that means the faster of the work of the legislative making the rule - will brings the governance system into the more efficient work of the rule applications. The small value of α made the system of governance to have over-damped response output. The over-damped response system has the steady-state in value of the state around 0.01 and either the system not oscillated.

In the other hand, the slower work of the rule application relative to the value of α - the work on the rule-making process- will make the system more efficient. In figure 5 (a and b), we can see that the system of governance will brings the steady state response approximately valued 1 just the same with the public want it to be while β=0.05. In this case we can see that the response will be oscillated at the first 8 time-sequence – while this also occurs in every value of β. It is apparent that the system in varied values of β made not as many oscillations like the varied values of α. Qualitatively, we can say that this occurs because α is the coefficient of the acceleration of rule-making process, the accentuation on the result of public inquiry, while β is the acceleration of how fast the rule is applied. Hence, we can say that the acceleration of the rule-application must "wait" the rule making process. The system of government response will be much more depend upon how the rule made. That is why the figure 5b is much smoother relative to the figure 4b.

Apparently, in the varied values of γ, while α=β=1 as we can see in figure 6, the more effective (high value of γ) the works of rule adjudication, the better the performance of the whole system. But something different here is that the better the rule adjudication, the system is not oscillated.

Qualitatively it is obvious. The work of rule-adjudication is to damp all of the disputes on the governmental system by arbitration. Henceforth, the better the rule-adjudication process, the system will be damped more. We can see this by the concave graph presented in figure 6b.

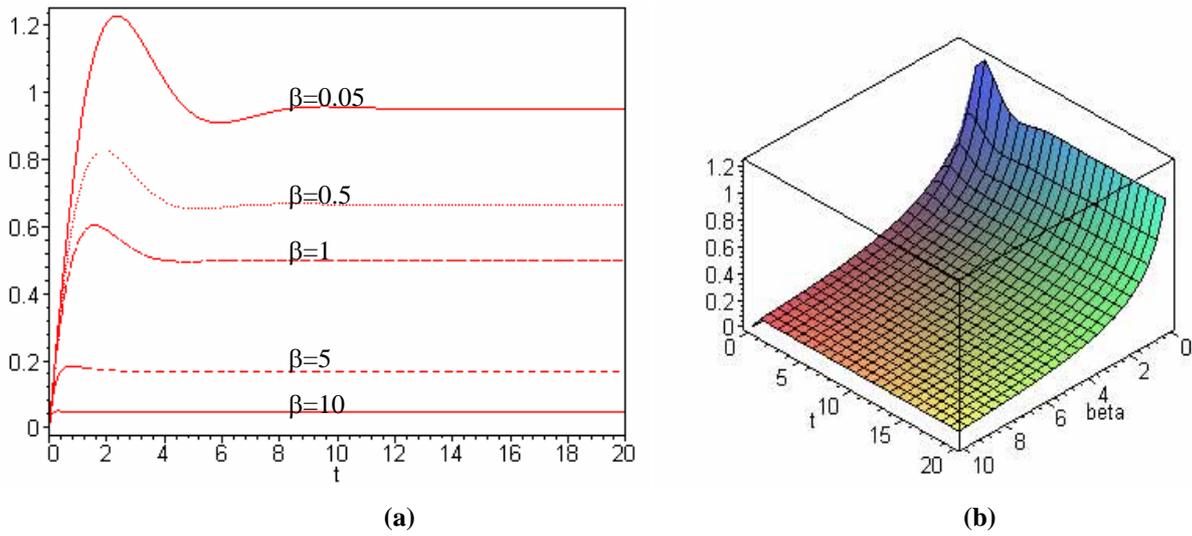

(a)                                              (b)

**Figure 5**
The response while α=γ=1 with varied values of β. The more value of the acceleration of execution or the application of the rule, the more efficient the works of the whole governmental system.

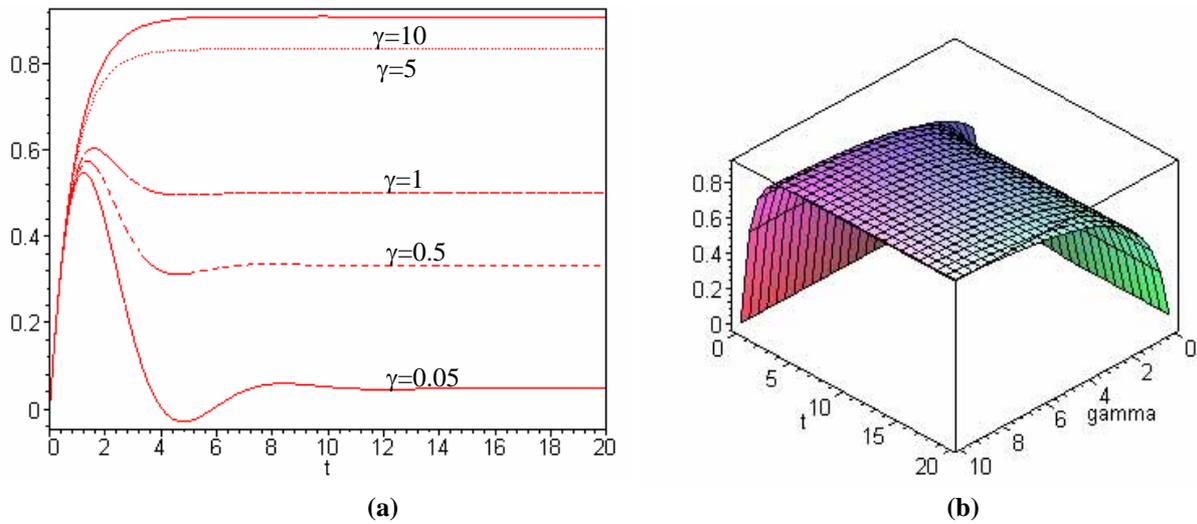

(a)                                              (b)

**Figure 6**
The response of the governmental system while α=β=1 in varied values of γ. It is obvious here the better the work of rule-adjudication, the system will be much more damped. The damped system of the disputes is projected by the damped system response.

### 4. The noise in the governmental system

The model we presented above is in the condition where there is no disturbance on the system of state that Montesquieu and Locke defined ideally. In some countries, there are some possible disturbances on the ideal system (Jacoby: 1985, Berg: 1987). For example, the possibility of military forces in the political area as we can say it in the military regime, the possibility of the corruption in the body of the state or even in the body of judicial elements, i.e.: judges, police officers, etc.

They are called the disturbance of the running government that ordered to serve the citizen. In our simulation, these possibilities can be assumed as a noise. A noise is the disturbance that bothers the elements of the system that even can bother the whole government system. Here, we use the band-limited white noise that generated randomly.

From the figure 7 we see how the government system disturb by such a noise that bother the legislative or the executive system. In control simulation, we place the noise generator before the rule-making or the rule-application function. By the figure we see that the response system become absurd, kind of chaotic response that reflected how the government will perform on the inability of the rule-making to aspirated the citizen's want or while the executive does not work properly. The signal presented in this condition will not make the public policies upon the public inquiries. Such a condition can turns in some despotic or dictator regimes. The use of the military forces in order to force the authoritarian or totalitarian regime can also projected such conditions (Arendt: 1960, p.392-393). The response of the system can still saw oscillated somehow but not oscillated in the fixed points that wanted by the citizen's wants (in the step response on value 1).

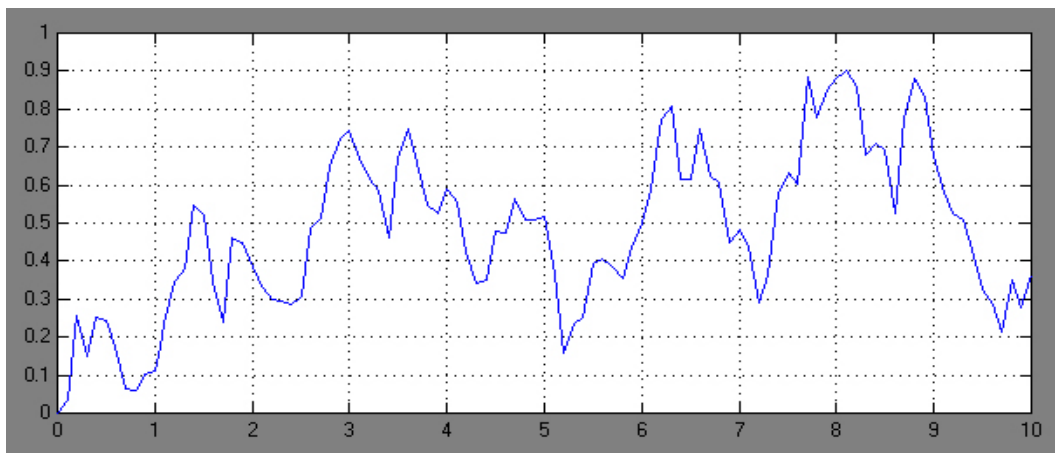

**Figure 7**
The governmental response function on the placement of noise at the forward path of the control dynamic system. This can reflected the absurdity of government that cannot projected the public wants.

The judicial system of government is the most importance elements in government, especially in the democratic system of governance (Bloch: 1986). The figure 8 shows the governmental system in the malfunction of the rule-adjudication. This is done by placing the limited-band white-noise at the backward path in the control system. We can see that the system become totally far from the how the citizens want it. The response system pictures that the governmental function become absurd so far henceforth the fail of the rule-adjudication function can be a fatal in a country: a country without justice in it.

However the corrupt dictatorship power in the state will gains the response system that much far from that the government should be. Figure 9 shows such a condition while all of the governmental elements are to be disturbed by the noise. The people can be forecasted to be stateless and henceforth cannot believe the government system. This is the utmost condition where the government is in a very far gap with the whole people. The response is chaotic and projects the situation where the whole states and citizens are restless. Probably, this is the condition where the governmental system in the despotic conditions as stated by Montesquieu once.

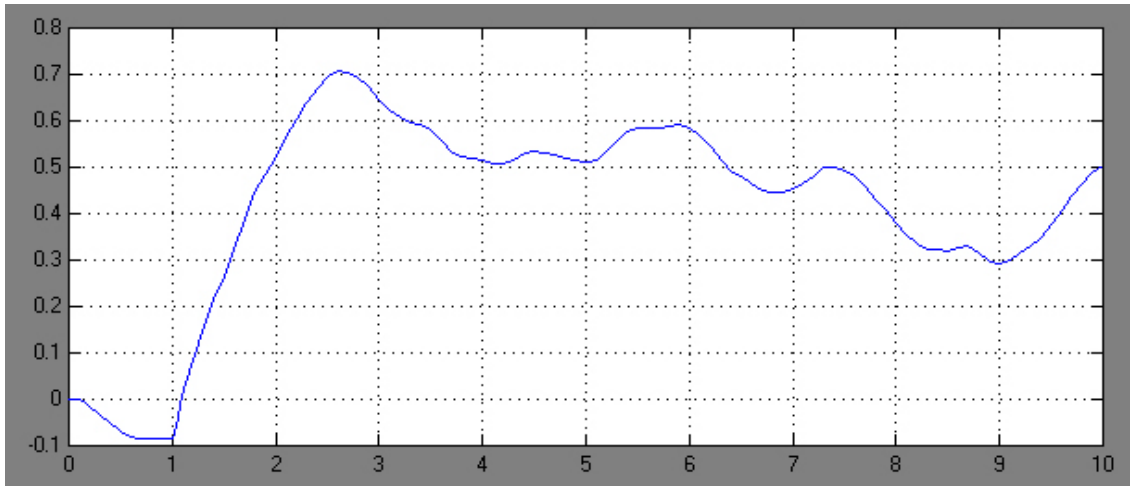

**Figure 8**
The governmental step response that turns out by placing the noise in the backward path of the control system.

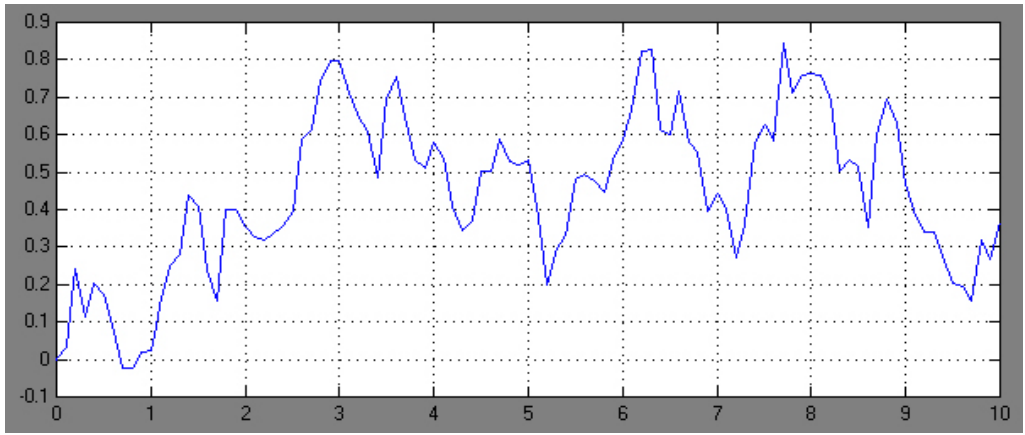

**Figure 9**
The response system where the whole elements of the government are disturbed by the noise become despotic. The chaotic signal reflects the restless conditions that appear in the whole state.

### 5. Further works

The simulations of the governmental model presented here are showed some conditions that so probable in our real world in many countries. However the paper shall be far away from the judgement of some countries that exists. Further works can be gained by practical research in some countries and made the scaling of the α, β, and γ therefore we can make a scientific judgement whether the governmental powers in a state have been good enough reflecting the citizens or not. It is obvious that the perfect response comes from the democratic system of governance as stated qualitatively by the classical works of Montesquieu and John Locke. But there will be no closed possibility that such a conditions can came from the government that not built from the democracy, but the functional elements i.e.: rule making rule adjudication, and rule application work in the proper way as simulated above.

## 6. Conclusion

Government system of a state can be seen as a feedback dynamic control system, where the rule-making and the rule-application function placed as plants in forward path of the system and the rule-adjudication placed in the backward path. The system of governance can be said to be reflected the citizens as long as the system can be optimized to have:

$$Lim_{t \to \infty} \frac{P_{in}(t)}{P_{out}(t)} \approx 1,$$

in a steady state response.

This can be achieved by setting the coefficients of each elements of the governmental system, i.e.: the coefficient of the acceleration of the rule-making process ($\alpha$), the coefficient of the rule-application function ($\beta$), and the coefficient of the rule adjudication function ($\gamma$). The value of the $\alpha$, $\beta$, and $\gamma$ shall be made such a way that the response system can gained damped solution in the steady state. While this situation cannot be achieved properly, the government system will be unstable and cannot reflect the citizens and badly can bring the society of the state to become restless public.

The despotic government system can come while the plants of the control system are disturbed by the noise that possibly can come from some negative aspects of the government. Qualitatively this can come from the corrupt elements, the authoritarianism using military forces, et cetera. However, the most important thing in the public area is the guarantee for justice. Henceforth, any disturbances in judicial elements of a state will turns the system in-comprehensively. While this is happen the system of governance will loose the teeth for the justice and can be very dangerous to the sustainability of the regime. The backward plants of the control system government must be set to have the fully damping any disputes that come from anywhere of the state. Rationally we can say that the most important thing in governmental system is the rule of the law in the state.

The democratic system of a state will brings a good response step in a nation, where the elements of the government will always be able to co-evolve such way to represents the citizens. The separation of the powers, minority rights, the civil society above the military forces, et cetera. However, we will not closed any possibility that a non-democratic system can gain such a good government that can be able represent the citizens. But still, the element functions of power must obey such conditions that presented in the simulation above.